\documentclass[prb,preprint]{revtex4-1} 

\usepackage{amsmath}
\usepackage{amsfonts} 
\usepackage{graphicx}

\begin{document}

\title{Computation and Experimentation as Equal Partners \\in a Modern Physics Lab Exercise}

\date{\today}

\author{Martha-Elizabeth Baylor}\email{mbaylor@carleton.edu.} 
\author{Luis A.~Miranda Almanzar}
\author{Adward Frazier Jr.}
\author{Adam Mahabir}
\author{Jay D.~Tasson}
\affiliation{Physics and Astronomy Department, Carleton College, Northfield, MN 55057}

\begin{abstract}
Experience with experimental and computational work are important components of students' understanding of the practice of physics. Physics departments typically use specific experimental lab courses and computational lab courses to develop students' skills in these areas. However these siloed experiences do not accurately represent the nuanced interplay between the theory, computation, and experiment that occurs outside of the curricular setting when physicists are building and testing models of the physical world. To expose students to this interplay, we have integrated computational and experimental work as equal partners within two radioactivity labs that occur within our modern physics lab course - the decay of $^{137}{\rm Ba}$ and the simultaneous decay of two silver isotopes in pre-1965 quarters. We present both of these labs and potential extensions that allow students to iterate on their theoretical, computational and/or experimental models based on what they learn from their initial investigation.
\end{abstract}

\maketitle

\section{Introduction} 

Undergraduate physics programs work to build student knowledge of theoretical and experimental techniques, with an increasing emphasis on incorporating computational techniques as well.\cite{general_computation} Faculty design courses to take conceptual knowledge and integrate that knowledge with techniques for solving problems in theoretical, experimental and computational approaches\cite{computation_litrev,experiment_litrev}. Courses that include laboratory exercises typically provide a space for students to practice combining theoretical and experimental OR theoretical and computational techniques in a pedagogically intentional environment. The goals of these types of laboratory experiences vary, but three primary goals include: teaching students discrete skills with applications to each area (e.g., theory, experiment, and computation) so that students develop competency in each, helping students decide which type of physicists they might want to become (e.g., theorist, experimentalist, computationalist), and helping students appreciate the types of questions and concerns in each of these approaches to doing physics.\cite{computation_litrev,experiment_litrev}

Unfortunately, the historical approach to teaching students theory, experiment, and computation does not expose them to the sophisticated interplay between physical systems, theory, experimentation, and computation that is becoming increasingly relevant.\cite{apsnews} As a result, students have a limited understanding of how physical models develop and evolve, particularly in collaborative environments between physicists with different types of expertise. Attending to the interplay between theory, experiment, and computation can provide students a more nuanced understanding of the “scientific method” and give them a more realistic understanding of doing physics outside of an academic setting. We develop single and multi-week labs that combine theory, experiment, and computation on equal footing to understand a physical system within a modern physics context.

In this paper, we first introduce our educational context and pedagogical goals. Next we introduce a series of integrated computational and experimental laboratory exercises focused on the decay of a single atomic species followed by a series of exercises focused on the simultaneous decay of two-atomic species. We conclude by discussing our observations performing these exercises over one or more lab periods. Additionally we highlight opportunities and challenges of our approach to these types of integrated experimental and computational laboratory exercises.

\section{Educational Context}
Carleton College is a private small liberal arts college located in Northfield, MN. Carleton is on a trimester schedule that consists of three 9.5-week terms with roughly 9 four-hour labs possible over the course of one trimester. The labs discussed in this paper are part of a 200-level Atomic and Nuclear Physics course (A\&N) that would be similar to a third-semester modern physics course in most institutions. A\&N is an introductory quantum mechanics course that includes applications to topics such as thermodynamics, solid state physics, and radioactive decay. Carleton requires one term of introductory physics (mechanics and relativity) before students are able to take 200-level courses and A\&N is typically the first course for majors that students take in our department, since the introductory courses do not separate majors from non-majors. A\&N typically enrolls 24-36 students, mostly physics majors and students that anticipate declaring physics as their major, split into lab sections of 12 students maximum with one instructor and one student lab assistant. Since students are not allowed to declare their major until their sixth term at Carleton, we aim to ensure that students have a positive experience in this course that is representative of the major so that genuinely interested students will declare physics as their major at the end of their sixth term. Students typically work in groups of 2-3 individuals (ideally 3 students), with opportunities for single week and multi-week labs. The labs require no prior exposure to programming, so an introduction to python code is included within the lab portion of the course.

\section{Background for Single-Atomic Decay}
\label{sec:SingleDecayBkgd}

\subsection{Ideal Radioactive Decay Model}
Most textbooks explain radioactive decay using Rutherford's and Soddy's model,\cite{rsmodel} which states that the rate at which a substance decays is constant. Mathematically, this is represented as
\begin{equation}
\frac{dN}{dt}=-\lambda N,
\label{eq:differential}
\end{equation}
where $\frac{dN}{dt}$ is the rate of change in the number of atoms, $N$, and $\lambda$ is the decay constant. Importantly, $\lambda$ is the probability that an atom will decay per unit time. Equation \ref{eq:differential} is the fundamental differential equation to nuclear decay of atoms. Students solve Eq. \ref{eq:differential} to find the amount of radioactive material as a function of time $t$, $N(t)$, as
\begin{equation}
 N(t)=N_0e^{-\lambda t},
\label{eq:Nt}
\end{equation}
where $N_0$ is the initial amount of the radioactive substance. From Eq.~\ref{eq:Nt}, we find the half-life, $t_{1/2}$, to be the time elapsed for half of the substance to decay: 
\begin{equation}
 t_{1/2}=\frac{\ln(2)}{\lambda}.
\label{eq:halflife}
\end{equation}
Equation \ref{eq:Nt} models the process as deterministic, as would be the case in the limit of an infinite number of atoms. They then proceed by linearizing Eq.~\ref{eq:Nt} as
\begin{equation}
\ln[N(t)]=\ln(N_0)-\lambda t.
\label{eq:linearized_N}
\end{equation}                                                                  
This straightforward analysis of radioactive decay does not take into account the fact that in an experimental context, students will not have a large enough sample of decaying atoms to ignore the probabilistic nature of the radioactive process. To consider the probabilistic nature of decay, we introduce students to the Monte Carlo computational technique to allow a more realistic representation of what they will likely observe in their experiment and to determine whether the linearization derived in the deterministic case (Eq. \ref{eq:linearized_N}) will allow them to recover the known half-life of an atom when probabilistic fluctuations are significant.  In this way, students can hone their data analysis techniques on simulated or ``mock" data before applying the technique to experimental data, as is common practice in research contexts.\cite{mockdata}

\subsection{Monte Carlo Technique Applied to a Single Isotope Decay}
We use the Monte Carlo technique to simulate the outcome of a random series of events that occur with some known probability. Examples of such systems include rolling dice or flipping a coin. When rolling a balanced six-sided die, the probability of any one side to be facing up is $1/6$. The action of rolling can be simulated with Monte Carlo by assigning each roll a random number between 0 and 1. If that value falls in the range $0<a<0.1\bar6$, then that roll is a face-up 1. Similarly, a face-up 2 is determined by a value of $0.1\bar6\leq b<0.\bar3$ and so on for all of the numbers one through six of a six-sided die. We represent a player rolling the die 10 times by using a random number generator to obtain 10 numbers between 0 and 1 as shown: 
\begin{equation}
[\:0.22,\;0.17,\;0.16,\;0.64,\;0.28,\;0.87,\;0.43,\;0.93,\;0.009,\;0.30\:].
\label{eq:RandomNumbers}
\end{equation}  
In the case of Eq.~\ref{eq:RandomNumbers}, the person rolled a `1' twice and a `2' four times.

For the case of simulating radioactive decay, we determine the probability $P$ of any atom decaying using
\begin{equation}
P=\lambda\, dt,
\label{eq:DecayProb}
\end{equation} where $dt<<t_{1/2}$ is the time step used in the simulation. To simulate whether any particular atom in a sample with $N$ radioactive atoms decays, we generate a list with $N$ random numbers between 0 and 1. If the number $a$ associated with the $i$-th atom $N_i$ is $0\leq a \leq \lambda t$, then we say that the $N_i$-th atom decayed and we remove it from the list of atoms. These decayed atoms can also be counted and recorded as the number of counts that would be measured by a detector.   

\section{Implementation of Single-Atomic Decay Lab}
\label{sec:SingleAtomImplement}
This section describes what students do during the four-hour lab period. The first half of lab focuses on using the Monte Carlo technique introduced in Sec. \ref{sec:SingleDecayBkgd}B to simulate the radioactive decay of an atom. The goal of this portion of the lab is to confirm that the linearization of the exponential decay from the theory continues to apply when the decay is not deterministic. During the second half of the lab, students observe the decay of an actual radioactive isotope and use the same analysis techniques that they verified with their simulated atoms to recover the half-life of the isotope. We discuss what happens in each portion of the lab in more detail below.

\subsection{Computational Implementation}
\begin{figure}[htbp]   
\centering
\includegraphics[width=16cm,keepaspectratio]{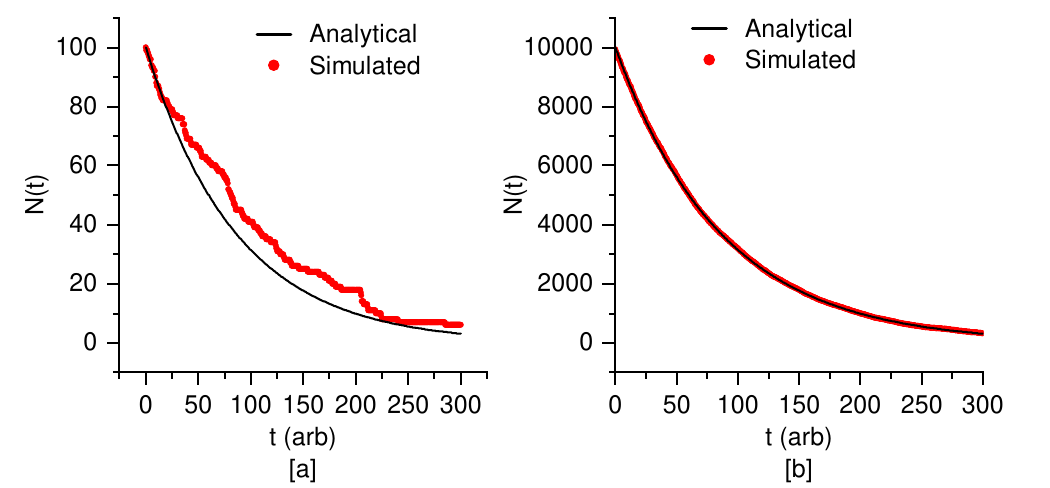}
\caption{Graphs showing number of atoms remaining $N(t)$ versus time $(t)$ using the ideal exponential decay equation (Eq. \ref{eq:Nt}) and a simulated computational implementation of radioactive decay using the Monte Carlo technique. The atoms in both simulations have $t_{1/2}=60$, but the initial number of atoms $N_0$ is varied. In [a], $N_0=100$, and in [b], $N_0=10000$.}
\label{fig:RemainingAtoms}
\end{figure}

The quantum mechanical introduction to radioactive decay occurs in the latter weeks of the course after students would have completed this experiment. So to orient students to single-atom atomic decay, students complete a brief textbook reading and pre-lab exercises that introduce them to the material summarized in Section \ref{sec:SingleDecayBkgd}. Students receive sample code that simulates and graphs the number of atoms remaining $N(t)$ as a function of time $t$ for an atom with a given $t_{1/2}$, where students understand that temporal variables are understood to have arbitrary units within the code. The computational model that students develop accurately shows the stochastic nature of radioactive decay (see Fig. \ref{fig:RemainingAtoms}a). Additionally, the simulation shows that as one increases the initial number of atoms $N_0$, the graph of simulated $N(t)$ vs $t$ shows less fluctuation and approaches the theorized deterministic curve (See Fig. \ref{fig:RemainingAtoms}b).

Students then modify the code so that it graphs the number of radioactive counts measured $C(t)$ as a function of $t$ to better align with what they would measure in an experiment. Note that the form of $C(t)$ matches the form of $N(t)$ such that
\begin{equation}
 C(t)=C_0e^{-\lambda t},
\label{eq:Ct}
\end{equation}
and the linearized form of $C(t)$ is
\begin{equation}
 \ln [C(t)]=\ln C_0-\lambda t.
\label{eq:linearized_C}
\end{equation}

\begin{figure}[htbp]   
\centering
\includegraphics[width=16cm,keepaspectratio]{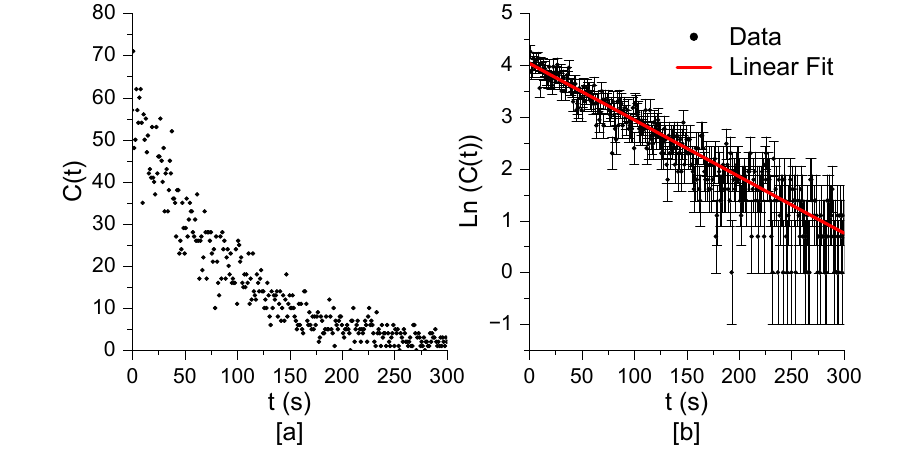}
\caption{Sample student simulated data, $C(t)$ and associated linearized $\ln [C(t)]$, using the parameters given in Eq.~\ref{eq:InitialConditions}. Subfigure [a] demonstrates that the `measured' counts follow the exponential decay in Eq.~\ref{eq:Ct}, while [b] shows that the $C(t)$ is linearized according to Eq.~\ref{eq:linearized_C} that gives the slope $-\lambda=-0.01092 \pm 0.00014 \;{\rm s}^{-1}$, which corresponds to $t_{1/2}=63 \pm 2 \;{\rm s}$.}
\label{fig:CountsVsTime}
\end{figure}

Students are encouraged to manipulate $N_0$, the time step ($dt$), the time duration of the experiment ($T$), and $t_{1/2}$ of the simulated atom to see the impact of these parameters on what they observe in their graph. After students observe the impact of changing these parameters, typically students return to the initial conditions they were provided in the original code that are found here:
\begin{equation}
N_0 = 10,000\; {\rm atoms}\qquad
t_{1/2}=60\; {\rm s}\qquad
T=300\; {\rm s}\qquad
dt=0.5\; {\rm s},
\label{eq:InitialConditions}
\end{equation}
where, as a matter of convenience, we choose seconds as the arbitrary unit of time that we will use for the remainder of the paper. We provide a graph of the simulated $C(t)$ versus $t$ using the initial conditions provided in Eq.~\ref{eq:InitialConditions} in Fig.~\ref{fig:CountsVsTime}a.

Students run the simulation and export the counts and the associated time to a spreadsheet. They then compute the uncertainty in time and counts ($\sqrt{C(t)}$) for each data point before linearizing $C(t)$ according to Eq.~\ref{eq:linearized_C}. Students plot $\ln[C(t)]$ versus $t$ including appropriate error bars, where the slope should be $-\lambda$.\footnote{The first time that $N(t)=0$, we removed the rest of the counts from the analysis.} 
Students observe that their data appears to be linear (see Fig. \ref{fig:CountsVsTime}b) with $\chi^2 \approx 1$ from weighted least squares linear regression. Students use the slope from the linear fit to determine $t_{1/2}$ for their simulated atom using Eq.~\ref{eq:halflife} and determine the uncertainty in this value using standard propagation of uncertainty techniques. Ideally, students should extract the $t_{1/2}$ value that they assigned their atom, confirming that their analysis procedure will allow them to extract $t_{1/2}$ of an unknown experimental sample.

We allow students to spend two hours exploring radioactive decay with their simulation, including analysing their computationally generated counts.
After students are reasonably convinced that the analysis procedure for the simulated data is effective, they measure the decay of a known, commercially available, and reasonably priced radioactive isotope, using a Geiger counter and Logger Pro software.

\subsection{Experimental implementation}
\begin{figure}[htbp]   
\centering
\includegraphics[width=10cm,keepaspectratio]{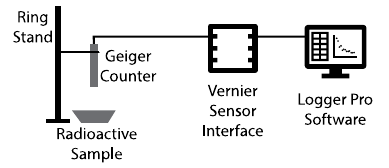}
\caption{Schematic of the experimental set-up. A ring stand holds the Geiger counter above the radioactive sample. The pulses from the Geiger counter are read by the Vernier Sensor Interface and the pulses are recorded and displayed on a computer using the Logger Pro software. Not shown is lead shielding around the ring stand, Geiger counter, and sample to limit radiation exposure.}
\label{fig:Schematic}
\end{figure}

For the latter two hours of lab, students measure and analyze the decay of $^{137}{\rm Ba}$.  We obtain the $^{137}{\rm Ba}$ material from the Spectrum Technologies Isotope Generator Kit available from Pasco (Part Number SN-7995A). The $t_{1/2}$ for $^{137}{\rm Ba}$ is $153\; {\rm s}$, but this information is not shared with the students until they have provided their best estimate of their experimentally measured $t_{1/2}$ for $^{137}{\rm Ba}$ . In the physical experiment, students can adjust the height of the Geiger counter above the sample, the amount and location of lead shielding, the amount of radioactive material via the number of drops of an elution process, the time bin size (i.e., the amount of time the Logger Pro measures counts for each data point), and the length of time Logger Pro measures the decay. Figure \ref{fig:Schematic} shows a schematic of the apparatus.  

Students are generally surprised by how few counts they receive when they initially take data and the resulting jumpiness between adjacent data points. This causes students to reflect on and adjust their experimental apparatus, sample preparation procedure, and Logger Pro Settings (e.g., the time step) to improve the quality of their data. Debates regarding the benefits of long versus short time steps and the need to find a balance that is appropriate for their physical situation allows for important discussions about the connection between their simulated decay and the actual decay students observe. Students perform the experiment multiple times to optimize their experimental data to reduce the stochastic fluctuations as much as possible when they record $C(t)$ from the Geiger counter as a function of elapsed time $t$.

Though students might ask about background radiation, we do not allow the students to consider background since their analysis is based on what they learned in the textbook that does not mention background radiation and, prior to doing the experiment, they have no observations of the existence of background radiation. We do encourage students to note anything they observe or other aspects of the experimental implementation that are not included in their physical model and/or their simulation.

Using the same spreadsheet that they used to perform the calculations for the simulated data, students determine uncertainty in their counts, calculate $\ln[C(t)]$, propagate uncertainty, and produce a plot of $\ln[C(t)]$ versus $t$ with error bars for their experimental data. Students recover an experimental $t_{1/2}$ for $^{137}{\rm Ba}$ from the slope of their linearized graph and compare their measured value to the expected $t_{1/2}$.

\subsection{Computational and Experimental Results}

Using the initial conditions in Eq.~\ref{eq:InitialConditions} for their computational code, students typically record a half-life that is close to, but does not include the expected half life within the uncertainty of the measured value. Moreover, the measured values are consistently higher than the expected value. For example for the simulated data shown in Fig.~\ref{fig:CountsVsTime}a, students recover $t_{1/2}=63 \pm 2 \;{\rm s}$ from the linear fit in Fig.~\ref{fig:CountsVsTime}b, which demonstrates both of the issues stated above. Since the discrepancy is small and the students typically perform the analysis on just one data set, these issues typically do not stand out to students and they are content to proceed to the experimental portion of the lab. Using the analysis procedure that they used on the simulated data on the $C(t)$ measured from the Geiger counter yields an experimental measurement of $t_{1/2}$ for $^{137}{\rm Ba}$ that is higher than the accepted value and precise enough that the uncertainty does not include the accepted value. Again, because the students only see their own results, they do not necessarily recognize that their computationally derived half-life for their test atom was also higher than the expected value and not within uncertainty.        

At this point, students are asked to consider what might be causing the discrepancy between the measured experimental value of $t_{1/2}$ and the accepted value. This prompts many students to think critically about their understanding of their apparatus and experimental process. Students mention the observation of counts before they place the radioactive sample below the Geiger counter. The presence of counts in the ambient environment is not included in their theoretical model or their simulation, but now students have observational evidence of a new source of counts that needs to be represented in their model. Students also frequently comment on the ways in which their simulation does not match the experiment. For example, the students assume many more decaying atoms in their simulation compared to the number of $^{137}{\rm Ba}$ atoms decaying in their sample as evidenced by the significantly fewer initial counts. 

Depending on the number of times students ran the simulation, some students also notice the slight bias that was present in recovering the half life from the simulation using the least squares fit to the linearized data.  This observation provides the opportunity to point again to the value of simulation and to consider the advantages and disadvantages of defaulting to linearization and least squares fitting.  Linearization provides visual feedback on the way the model fits the data, and allows a good (though not perfect) analysis without the introduction of the more sophisticated techniques that improve the recovery of the half life.  Students with a background in statistics may reflect on issues of transformation bias in the linearization and the underlying probability distribution for the number of decay counts observed at a given time.\footnote{These ideas are typically more sophisticated than what most of our students are ready to handle at the beginning of their second year at Carleton.}

\section{Discussion of Computational and Experimental Implementation Single-Atomic Decay Lab}
\label{sec:discsingle}
The combined computational and experimental exploration of a single atom decay allows the instructor to foster deeper conversations around the relationship between theory, computation and experiment. Given the observation that neither the computational nor the experimental analyses produce the accepted value within uncertainty causes us to reflect on the purpose of these exercises. What is the purpose of measuring the half-life of a known, natural quantity if not to recover the accepted value within uncertainty? Should we have continued to work on the computational model when our analysis did not result in the computational half-life? Certainly the constraints of the length of the lab period influence our pedagogical choices. In this section, we consider additional changes or modifications to what we presented in Section \ref{sec:SingleDecayBkgd}. If it were possible to add an additional lab period to the single isotope decay lab, we have several paths that we might pursue. 

\subsection{Modifying the Model to account for Background Radiation}
\label{sec:discsingle_bkgd}
One approach would be to ask students to go back to their theoretical model and add the physics that is missing (i.e., the observation of background radiation) based on their experimental observation. Students could then take a more systematic look at the background radiation to determine whether it appears constant or random and modify their computational model and analysis approach before retrying their experimental work. We now discuss a separate exploration  of this path.

We observe that the background radiation counts are random, but the background count rate appears to be constant. From this point, we must differentiate counts from different sources. Therefore, we use superscript B to indicated quantities related to the background radiation and superscript T to indicated quantities related to total radiation measured at the detector. Since $C^{\rm B}$, the number of background counts, occurs at a rate that is constant up to random fluctuations, we modify our model to reflect the total counts at our detector, $C^{\rm T}(t)$, as
\begin{equation}
C^{\rm T}(t)=C_0 e^{-\lambda t} + C^{\rm B}.
\label{eq:CountRateDet}
\end{equation}
We linearize Eq.~\ref{eq:CountRateDet} as
\begin{equation}
\ln[C^{\rm T}(t)-C^{\rm B}]=\ln (C_0) -\lambda t.
\label{eq:LinCountRateDet}
\end{equation}
Thus, we preserve our intuitive sense that we should subtract the background counts from the measured counts to recover the counts of the isotope of interest. We obtain the half-life of our atom from the slope of a $\ln[C^{\rm T}(t)-C^{\rm B}]$ versus $t$ using Eq.~\ref{eq:halflife}.

To implement the new model represented by Eq. \ref{eq:CountRateDet} in our simulation, we reflect on what the background radiation could be. Background radiation includes many different, unrelated, physical processes (e.g., cosmic rays, isotopes in the local environment\footnote{An example might be concrete used in the building or the dirt in the region}, isotopes transported through the local environment\footnote{An example might be someone with a banana in their backpack or fiesta ware.}, etc.). So simulating background radiation accurately can be a challenge.  To reflect that the background count rate appears to be constant as specified in our model, we simulate background radiation as the radioactive decay of a non-diminishing population of an isotope.  The half-life and population size are chosen to reasonably simulate the very small (e.g., 1-2 counts every 3-5 seconds) background count rate that is typically observed in the lab.
Hence the background can be reasonably simulated by reusing the same code used in simulating the isotopes of interest with minor modifications.  

In an actual experiment, students would measure the background radiation separately from their sample decay, which includes both the decay from the isotope of interest and a background radiation. The background radiation that is measured with the sample is uncorrelated to the background radiation measured on its own. Thus, students generate and export two separate data files. The detector counts $C^{\rm T}(t)$ file contains the sum of the simulated atoms counts and background counts as a function of time. It is important that these counts are not differentiated into separate columns, since in an actual experiment, students do not know how many of their counts are from the atom of interest and which are from background. The second file contains only background counts as a function of time, making sure that the background counts are generated separately (i.e., are not the same values) as the background counts added to the simulated atom decay. In performing the analysis of the data using the model provided by Eq.~\ref{eq:LinCountRateDet}, students require $C^B$, which they do not have direct access to. Students instead approximate $C^B$ by calculating an average number of background counts
\begin{equation}
C^{\rm B}_{\rm avg}=\frac{dt}{T^{\rm B}}\sum C^{\rm B},
\label{eq:CalcRates}
\end{equation}
where $dt$ is the time step and $\sum C^{\rm B}$ is the sum of all of the background counts observed during the total time $T^{\rm B}$ over which we measure the background.

As suggested previously, the shift in analysis required using this new model represented by Eq.~\ref{eq:CountRateDet} impacts students' experimental approach, specifically how they measure the background counts. This approach nicely shows the way that experimental observation can feedback to theoretical and computational work to develop a more complete physical model (and therefore understanding) of a complex system. Forcing students to think about how they might model the background radiation in their simulation could result in further experiments and/or a literature search to determine the nature of the background radiation and how they might model this type of radiation.

\subsection{Alternative Methods for Recovering the Half-Life}
\label{sec:poisson}
\begin{figure}
    \centering
    \includegraphics[width=120mm]{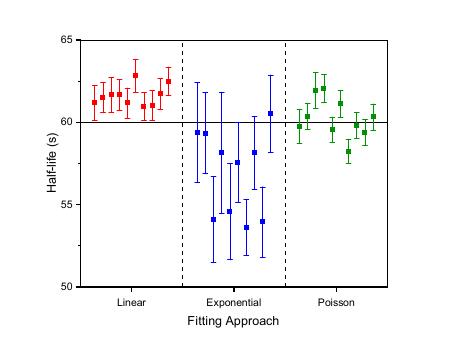}
    \caption{Graph of recovered half-lives using different fitting and regression techniques. We applied linear regression, exponential fit, and Poisson regression to the same 10 data sets simulating an atom with a 60 second half-life. The linear regression consistently produces a half-life that is larger than the expected half-life with an uncertainty that does not include the accepted value. A direct exponential fit to the data, produces a half-life that is shorter than the accepted value in most cases. The large error bars occasionally cause the accepted value to be recovered within uncertainty. The Poisson regression appropriately produces a half-life that appears randomly distributed about the accepted value and often includes the accepted value.}
    \label{fig:simulation_fit}
\end{figure}

Another path that we could pursue would be to ask students to grapple more deeply with their analysis approach when the application of weighted least squares regression to the linearized computational data fails to recover, within uncertainty, the half-life that they set in the simulation.  They would first need to explore the result of the simulation over a larger number of trials as shown in Fig.~\ref{fig:simulation_fit} to be convinced that a significant discrepancy exists.
There are then two related issues with the process of linearizing this data and applying weighted least squares fitting that can be considered.  We discuss each in turn.

First, the process of transforming data, doing statistical analysis on the transformed data, then applying the inverse transformation to the resulting statistical measures has the potential to yield biased results.  This occurs here in our linearized analysis as we take the natural log of the counts, fit the data, then exponentiate the fit parameters. In the context of curve fitting, this effect is known as transformation bias.  While a complete treatment of transformation bias is beyond our present scope, we point interested readers to the examples and discussion in Ref. \onlinecite{transf_bias}, and we offer the following toy example that illustrates the potential issue.

Consider a variable $x$.  If we take its natural log, then exponentiate the result, we recover $x$: $e^{\ln x}=x$.  However, if we consider a set of random variables $x_i$, compute the natural log of each, compute the mean, then exponentiate the result, we are not guaranteed to get the mean of $x$: $e^{\langle\ln(x)\rangle} \leq \langle x \rangle$.  One might imagine that this is the bias we see in 
Fig.~\ref{fig:simulation_fit}.  One might further expect that fitting directly to the exponential using nonlinear least squares fitting might address the issue.  However, Fig.~\ref{fig:simulation_fit} demonstrates that this approach does not lead to an entirely satisfactory solution. 

The second issue with the fitting approach is that least squares fitting (linear or nonlinear) is not exactly applicable to our data given that the observed decay counts at a given time are Poisson-distributed rather than Gaussian-distributed.  In discussing the issue with students, it may be useful to point out features such as the impossibility of negative counts and the fact that very high numbers of counts are possible but very rare.  Both of these features are properties of the Poisson distribution, but the former is at odds with the Gaussian distribution.  The fit can be performed correctly with Poisson regression.  This approach avoids transforming the data and uses the correct distribution.

To better understand the issue with the least squares fit and to understand Poisson regression, we turn to the foundations of fitting.  Suppose we have a set of measured data in the form of $N$ ordered pairs $\{(x_1,y_1\pm\sigma_1),(x_2,y_2\pm\sigma_2),...,(x_N,y_N\pm\sigma_N)\}$.  
In the present context, the roles of $x_i, y_i$ are played as follows: $x_i \rightarrow t_i$ and $y_i \rightarrow \ln C_i$, where we introduce $C_i=C(t_i)$ for simplicity. 
Further suppose that we have a model for this data that has two parameters, call them $A$ and $B$.  An example of such a model would be a linear model $\mu_i=A+B x_i$.  Here $\mu_i$ is the expectation value for the $y$ parameter at $x_i$.  
In the present context, $A\rightarrow \ln C_0$, $B \rightarrow -\lambda$, and $\mu_i$ is the expectation value of $\ln C_i$. 
If the possible $y$ values that may be observed at $x_i$ are Gaussian distributed, $\mu_i$ can be thought of as the mean of this distribution.  In general, we can write the probability of observing a particular value $y_i$ at $x_i$ given a particular pair of fit parameters $A$ and $B$ as $P_{A,B}(y_i)$.  In the Gaussian case, this would be 
\begin{equation}
P_{A,B}(y_i) = \frac{1}{\sigma_i \sqrt{2 \pi}} e^{-\frac{1}{2\sigma_i^2}(y_i - \mu_i)^2}.
\end{equation}
We then consider the joint probability of obtaining all of our observed data points given the model:
\begin{equation}
    P_{A,B}(y_1,y_2,...,y_N)=P_{A,B}(y_1) P_{A,B}(y_2)...P_{A,B}(y_N),
\end{equation}
which is the product of the probabilities to observe each data point.  The fitting process is then completed by employing the Principle of Maximum Likelihood by seeking the values of $A$ and $B$ that would make our data most likely, that is, the values that maximize $P_{A,B}(y_1,y_2,...,y_N)$.

Poisson regression is a special case of a class of models called generalized linear regression in which the fit equation may be a function $g$ (known as the link function) of our linear equation.  Hence, the expectation value of $y$ at a given $x_i$ could be written $\mu_i = g(A + B x_i)$.  
In the Poisson regression case, the link function is the exponential function and the data is assumed to be Poisson-distributed.  
So our probability of a particular $y_i$ at $x_i$ would now be given by the Poisson distribution 
\begin{equation}
P_{A,B}(y_i) = \frac{\mu_i^{y_i}e^{-\mu_i}}{y_i!},
\end{equation} 
where the $\mu_i$
are now provided by our link function.  
Notice that in our context, the roles of all parameters are the same with the exception of $y_i \rightarrow C_i$, and $\mu_i$, which is now the expectation value of $C_i$.
We may then proceed as before by applying the Principle of Maximum Likelihood to obtain values for our fit parameters $A$ and $B$.  In doing so, we'll now have avoided the transformation because we are fitting directly to the $C_i$ data rather than its natural log, and we'll have used the correct distribution, the Poisson distribution, for our counting data.  The applications of Poisson regression to our simulated data in Fig.~\ref{fig:simulation_fit} indeed appear to recover the input half life in a way that is consistent with expectations. 

\begin{figure}[htbp]   
\centering
\includegraphics[width=16cm,keepaspectratio]{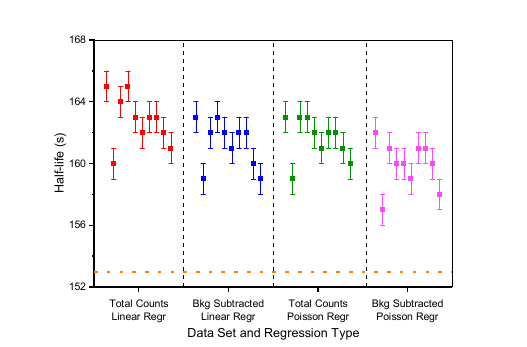}
\caption{Sample student experimental measurements of the $^{137}{\rm Ba}$ half life. The results from the analysis of the same 10 data sets using four fitting approaches are shown.  The accepted half life of $^{137}{\rm Ba}$ is indicated by the horizontal dashed line.}
\label{fig:expData}
\end{figure}

The impacts of both advanced fitting techniques and background subtraction on 10 student-generated experimental $^{137}{\rm Ba}$-decay data sets can be seen in Fig.~\ref{fig:expData}.
As expected,
the use of Poisson regression and the subtraction of background reduce the bias in the experimental half life; however, an additional bias still remains.
Hence students could continue to explore sources of bias in the experimental data that lie beyond the subtraction of background radiation in the room.  These additional experimental sources of bias could include detector dead time and bleed-through of $^{137}{\rm Cs}$ during the preparation of the sample.  Detector dead time could mean that counts are systematically depressed in proportion to the count rate.  Bleed-through of the very much longer lived $^{137}{\rm Cs}$ is effectively an additional source of background that is mixed with the sample.  It seems conceivable that both of these effects could bias results toward longer half lives.  Students could also consider more advanced fitting techniques that fit to the background-containing model for counts as a function of time.  

Consideration of background, advanced fitting techniques, and other sources of error offer rich opportunities to engage more deeply with the ideas of theory, computation, analysis, and experiment in a seemingly simple physical system, where they can reasonably iterate between these different modes of exploration and let these modes of exploration inform each other. Currently within the Carleton curriculum, we use the single atomic decay to serve as a knowledge base in computational modeling and data analysis of atomic decay to prepare students for a more complicated physical system of the simultaneous decay of two isotopes with different half-lives that cannot be physically separated for independent decay measurements. Therefore, we would reserve this line of discussion and/or exploration for interested or more advanced students.

\section{Background for Two-Atom Decay}

\subsection{Theoretical Model of Two-Atom Decay}
 For a system consisting of two isotopes decaying simultaneously in the presence of a constant background signal, we can model the measured counts as
\begin{equation}
C^{\rm T}(t)=C^{\rm L}(0) e^{-\lambda^{\rm L} t} + C^{\rm S}(0) e^{-\lambda^{\rm S} t} + C^{\rm B},
\label{eq:Two-AtomDecayCountRate}
\end{equation}
where we assumed that the decay constants are not equal such that one isotope will have a shorter $t_{1/2}$ represented by superscript S and the other isotope will have a longer $t_{1/2}$ represented by a superscript L. Note that this means that $\lambda^{\rm S}>\lambda^{\rm L}$. As with the single decay, we can subtract $C^{\rm B}$ from both sides of Eq.~\ref{eq:Two-AtomDecayCountRate} to obtain
\begin{equation}
C^{\rm T}(t)-C^{\rm B}=C^{\rm L}(0) e^{-\lambda^{\rm L} t} + C^{\rm S}(0) e^{-\lambda^{\rm S} t}.
\label{eq:Two-AtomDecayCountRate_SubBkgd}
\end{equation}
The next step would be to linearize Eq.~\ref{eq:Two-AtomDecayCountRate_SubBkgd} by taking the natural log, as we did with single-atom decay; however, this approach to linearization proves impossible because of the two distinct half-lives. Instead we notice that after a sufficiently long time, the counts due to the fast decaying species (i.e., the second term on the right in Eq.~\ref{eq:Two-AtomDecayCountRate_SubBkgd} with $S$) are negligible, and therefore Eq.~\ref{eq:Two-AtomDecayCountRate_SubBkgd} reduces to single isotope case (see Eq.~\ref{eq:CountRateDet}) that we linearize as Eq.~\ref{eq:LinCountRateDet}.

When they plot the natural log of Eq.~\ref{eq:Two-AtomDecayCountRate}, students may observe several features in the plot and residuals that  help us measure the half lives of the two isotopes. There are three phases seen in the plot, one where the fast-decaying isotope is dominant, one where both are essentially equal, and one where the slow decaying isotope is dominant. During the decay of the fast-decaying isotope, the contribution of the slow-decaying isotope is negligible, resulting in a linear trend. In the intermediate phase, the contributions from both isotopes balance, leading to a non-linear decay pattern. Finally, as the fast-decaying isotope diminishes to near-zero levels, the slow-decaying isotope becomes dominant, resulting in another linear decay pattern. Analyzing the rates of decay for both isotopes individually, as well as their combined total, reveals these distinct linear trends, alongside the overall non-linear decay pattern. We are able to create a method of analysis that allows us to recover the decay constant, and eventually measure the half lives of the two isotopes.  

\subsection{Monte Carlo and Two-Atom Decay}
The Monte Carlo technique discussed in Section \ref{sec:SingleDecayBkgd} applies to the two-atom decay as well. Based on the theoretical model represented by Eq.~\ref{eq:Two-AtomDecayCountRate}, there are three sources that students can use the Monte Carlo technique to simulate - the two atomic species with different half-lives and the background radiation. To model the two atomic species, each species is represented by its own application of the Monte Carlo technique. Practically this means using separate arrays to hold the information about the number of decayed atoms for each simulated isotope with $\lambda^{\rm L}$ and $\lambda^{\rm S}$. We can also include background radiation in this simulation as discussed in Sec.~\ref{sec:discsingle_bkgd}.

\section{Implementation of Two-Atom Decay Lab}
In the two-atom decay lab, students determine the half-lives of two different silver isotopes produced from quarters made before 1965. These quarters are composed of an alloy of 90\% silver and 10\% copper. The silver in the quarters is a mixture of approximately 53\% $^{107}{\rm Ag}$ and 47\% $^{109}{\rm Ag}$, which are both stable. However, when we irradiate $^{107}{\rm Ag}$ and $^{109}{\rm Ag}$ in a neutron howitzer, the Ag nuclei undergo neutron capture which converts them into unstable isotopes $^{108}{\rm Ag}$ and $^{110}{\rm Ag}$ that have half-lives of $t_{1/2}^{108}=145.2$~s and $t_{1/2}^{110}=24.6$~s, respectively.\cite{halflives} Thus $^{108}{\rm Ag}$ is the long half-life isotope to be identified with the L superscripts, while $^{110}{\rm Ag}$ is the short half-life isotope associated with the S superscripts.

Prior to the two-atom decay lab, our students complete the single-atom decay lab as described in Section \ref{sec:SingleAtomImplement}. The two-atom decay lab builds on the following skills introduced in the single-atom decay lab: how to use the Monte Carlo technique to simulate the decay of a single atomic species, how to use the simulated data to test their analysis approach, and how to optimize their experimental set-up to measure radioactive decay to minimize uncertainties.

Similar to the single atom decay, this experiment has two parts: a computational component to determine an analysis procedure to recover the half-life of simultaneously decaying atomic species, and an experimental component to apply their procedure to an actual physical system. At the conclusion of this two-day lab, students provide their best estimate of $t_{1/2}^{\rm L}$ and $t_{1/2}^{\rm S}$ with uncertainty.

\subsection{Computational Implementation}
On the first lab day, students develop a computational model that provides a realistic implementation of the two-atom decay model. Students visualize their simulation data to develop an analysis procedure to recover the half-lives of the individual decaying species and practice implementing their procedure on exported simulation data to verify the efficacy of their analysis protocol. The goal of the first lab day is to develop a \textit{robust} procedure that will allow them to determine the half-lives of two atomic species decaying simultaneously.

At the beginning of the lab session, students modify their code to apply the Monte Carlo technique to two different atomic species with different half lives. We initially ignore background for simplicity. The overall structure of the code remains the same. Students duplicate many of the variables, except for $t$ that is common to both decays, modifying variable names to keep the counts associated with each species separate for each time step. The total counts $C^{\rm T}(t)$ is the sum of the counts from the short $C^{\rm S}(t)$, long $C^{\rm L}(t)$ lived isotopes, and the counts from the background $C^{\rm B}(t)$ such that
\begin{equation}
C^{\rm T}(t)=C^{\rm S}(t)+C^{\rm L}(t)+C^{\rm B}(t),
\label{eq:TotalCounts2Atom}
\end{equation}
where $C^{\rm B}(t)=0$ for students' initial investigations. Students then simulate $C^{\rm S}(t)$ and $C^{\rm L}(t)$ using the following parameters:
\begin{equation}
N_1 = N_2 = 10^4\; {\rm atoms}\qquad
t_{1/2}^{\rm L}=60\; {\rm s}\qquad
t_{1/2}^{\rm S}=8\; {\rm s}\qquad
t_{\rm final}=300\; {\rm s}\qquad
dt=0.5\; {\rm s}.
\label{eq:InitialConditions2Atom}
\end{equation}

\begin{figure}[htbp]   
\centering
\includegraphics[width=16cm,keepaspectratio]{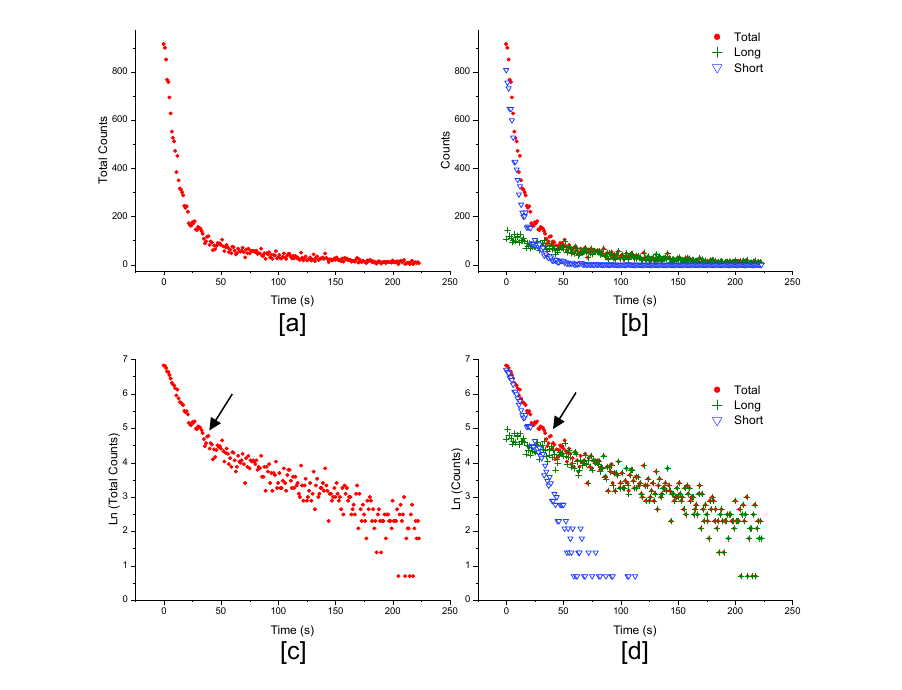}
\caption{Sample student simulated data for two atom decay without background. [a] Shows total counts $C^{\rm T}(t)$ versus time $t$ for a two-atom decay. [b] Shows the $C^{\rm T}(t)$ versus $t$ and the counts associated fast and slow decaying species, $C^{\rm S}(t)$ and $C^{\rm L}(t)$ respectively, that contribute to $C^{\rm T}(t)$. Unfortunately the relationship between these three data sets are difficult to observe. [c] Shows $\ln[C^{\rm T}(t)]$ vs $t$ that clearly displays two linear regions, one linear region for $t<25$~s and one linear region for $t>25$~s. The transition between these two regions is called `the knee' and is indicated by the arrow. [d] By plotting $\ln[C^{\rm S}(t)]$ and $\ln[C^{\rm L}(t)]$ alongside $C^{\rm T}(t)$, students observe that the fast decay dominates for $t<20$s and that for $t>120$s, $C^{\rm T}(t)=C^{\rm L}(t)$. These observations allow student to develop an analysis approach to recover $t_{1/2}^{\rm L}$ and $t_{1/2}^{\rm S}$.}
\label{fig:Sim4PanelCounts}
\end{figure}

Students are reminded that in the experimental portion of the work, they will only be able to observe the total number of counts $C^T(t)$, and will be unable to distinguish which counts are coming from which source
(see Fig.~\ref{fig:Sim4PanelCounts}a). On the other hand, the simulation allows students to plot the counts versus time for each isotope and the total counts separately (see Fig.~\ref{fig:Sim4PanelCounts}b), color-coded for clarity. Figure \ref{fig:Sim4PanelCounts}b allows students to observe characteristic aspects of the total counts due to features of the individual component counts and determine how best to approach disentangling the counts to recover the half-lives of the desired isotopes. Unfortunately, observing $C^{\rm T}(t)$, $C^{\rm S}(t)$, and $C^{\rm L}(t)$ versus $t$ on a linear scale does not provide the desired clarity.  
We prompt the student to plot $\ln[C^{\rm T}(t)]$, even though they are aware that this operation will not linearize their data. Students observe two approximately linear regions with different slopes, an observation that clarifies the purpose of plotting the natural log at this stage.  At the intersection between of these two slopes, students observe  a ``kink" or``knee" in the graph (see Fig.~\ref{fig:Sim4PanelCounts}c). This is an opportunity for students to think about what might be going on to cause this ``knee".

Next we prompt students to plot $\ln[C^{\rm S}(t)]$, $\ln[C^{\rm L}(t)]$, and $\ln[C^{\rm T}(t)]$ on the same color-coded graph (see Fig. \ref{fig:Sim4PanelCounts}d). Figure \ref{fig:Sim4PanelCounts}d demonstrates that for $t\lesssim25$~s, $C^{\rm S}(t)$ dominates. The knee occurs in the range $25s\lesssim t \lesssim 50$~s. At a sufficient time after the knee, in this case $t \gtrsim 125$~s, the fast isotope has decayed such that $C^{\rm T}(t)=C^{\rm L}(t)$ to an excellent approximation. Thus if students select some appropriate time after the knee, they can linearize or otherwise fit their data to find $\lambda^{\rm L}$ and $C_0^{\rm L}$ for their slow isotope from the fit parameters. Once students have recovered $\lambda^{\rm L}$ and $C_0^{\rm L}$, they can extrapolate $C^{\rm L}(t)$ for $t\lesssim25$~s where $C^{\rm S}(t)$ dominates and remove these counts from the total counts, isolating $C^{\rm S}(t)$. Once students isolate $C^{\rm S}(t)$ for $t\lesssim25$~s, they can linearize or otherwise fit these data to recover the fit parameters $\lambda^{\rm S}$ and $C_0^{\rm S}$ for $C^{\rm S}(t)$.

Once students reason through the general procedure described in the previous paragraph, they vary $t_{1/2}^{\rm L}$ and $t_{1/2}^{\rm S}$ so that they are closer and further apart. They reproduce Fig. \ref{fig:Sim4PanelCounts}d for several cases of their choosing and observe the visibility of the knee and the underlying relationships between $C^{\rm S}(t)$, $C^{\rm L}(t)$, and $C^{\rm T}(t)$.  We encourage students to develop a robust procedure for identifying the knee, the time cutoff above which they will assume $C^{\rm T}(t)=C^{\rm L}(t)$ and the time cutoff below which they will assume that $C^{\rm S}(t)$ dominates. At this point, we also ask students to consider how they will deal with background radiation in their procedure, building from their observations during the single isotope lab. Students finalize their simulation by adding the model for background radiation into their code.

At the conclusion of the first lab day, each group explains their procedure to the instructor for dealing with background radiation and recovering the half-lives of their two simulated isotopes. The instructor asks probing questions to ensure students understand what they are doing and why. Each group generates two data files. One data file corresponds to the data that students would receive at their Geiger counter and has two columns, $t$ and $C^{\rm T}(t)=C^{\rm S}(t)+C^{\rm L}(t)+C^{\rm B}(t)$. The second file corresponds to the background radiation measurement and has two columns $t$ and $C^{\rm B}(t)$. Students must re-run their code that generates the counts for the background radiation file so that the background radiation included as part of $C^{\rm T}(t)$ is not correlated with $C^{\rm B}(t)$ in their background file. This reinforces the reason students cannot perform a point-by-point direct subtraction of background counts from their total counts. In preparation for the second lab session, each group member independently implements their group procedure and recovers a half-life with uncertainty for each isotope.

\subsection{Experimental Implementation}
At the start of the second lab day, students meet with their lab group and each group member shares the half-lives with uncertainty that they determined during their individual analysis. Since students are not supposed to talk to each other when they implement their analysis procedure on the simulated data, it is common that they do not identify the knee to be in the same place and they choose different time ranges to analyze the fast and slow decay. These choices are subjective and students are not typically precise in their analysis procedure in how they will make these choices. We hope that by sharing their individual results and discussing which choices resulted in better agreement with the known simulated atom half-lives that each group will be able to further refine their analysis procedure before working with experimental data. This discussion takes approximately 15 minutes before students move to working on the experiment.

The experimental portion of the lab requires students to irradiate pre-1965 quarters with a neutron Howitzer.\cite{howitzer} We recommend that students irradiate the quarters for a minimum of 10 minutes, however this is an experimental parameter that students can vary. The time to transfer the quarters is extremely important given how few counts students are likely to measure and the short half-life of the fast decay. Students practice transferring un-radiated quarters from the neutron Howitzer to the Geiger counter as quickly as possible, preferable under 7 s.\footnote{We arrange the lab so that students can place their measurement apparatus as close to the neutron Howitzer as is reasonable with a straight running path to their apparatus. Students specifically wear running shoes and thoroughly enjoy this active portion of the lab.}

Deciding the experimental settings requires careful thought and students are encouraged to use their observations from the single-atomic decay to inform choices they make in this experiment. Though we provide students information about the composition of the quarters, we do not provide students with the half-lives of the two isotopes and we ask the students not to look up the half-lives. Rather, the instructor informs students that the shorter-lived isotope has $t_{1/2}^{\rm S}<30$~s and the longer-lived isotope has $100 \; {\rm s}<t_{1/2}^{\rm L}<200$~s, and suggests a time bin of 0.1 s - 4 s. Additionally, based on their observation of background radiation from the single atom decay, students decide how they will measure their background radiation in a way that is consistent with their simulation. We ask students to consider the following questions as they determine their settings:
\begin{itemize}
  \item Does it matter if they start taking data before the quarters have reached the detector?
  \item Does the time between when the quarters are irradiated and when the quarters reach the detector matter?
  \item What happens if the time bin size is too short or too long?
\item Does it matter if they measure background for the same experiment duration and time step as what they use for the quarters?
\end{itemize}
Once each group decides on their initial settings, the instructor informs the students of the radiation safety protocol the class will use when operating the Howitzer to minimize radiation exposure. In short, students are not allowed to enter the room with the neutron Howitzer when the neutron source is raised to expose the quarters to the plutonium-beryllium source. Only after the instructor lowers the source back into the shielding are students allowed to retrieve their quarters to place under the Geiger counter.

Given that students developed the spreadsheets to analyze their simulated data, they are typically able to quickly analyze their experimental data in the same fashion. This allows students to see their initial data and then modify their experimental parameters to improve the quality of their data (i.e., more counts in further trials). Students also further reflect on their analysis procedure, which can also impact decisions they make about their experimental parameters. Students are typically able to analyze 2-4 datasets during our 4-hour lab periods. Once the students settle on their experimentally derived half-lives for the two silver isotopes, they share these values and the associated uncertainty with the instructor, who then provides the experimentally accepted values of the half-lives of the silver isotopes. 

\subsection{Discussion of Two-Atomic Decay Lab Implementation}
\begin{figure}[htbp]   
\centering
\includegraphics[width=16cm,keepaspectratio]{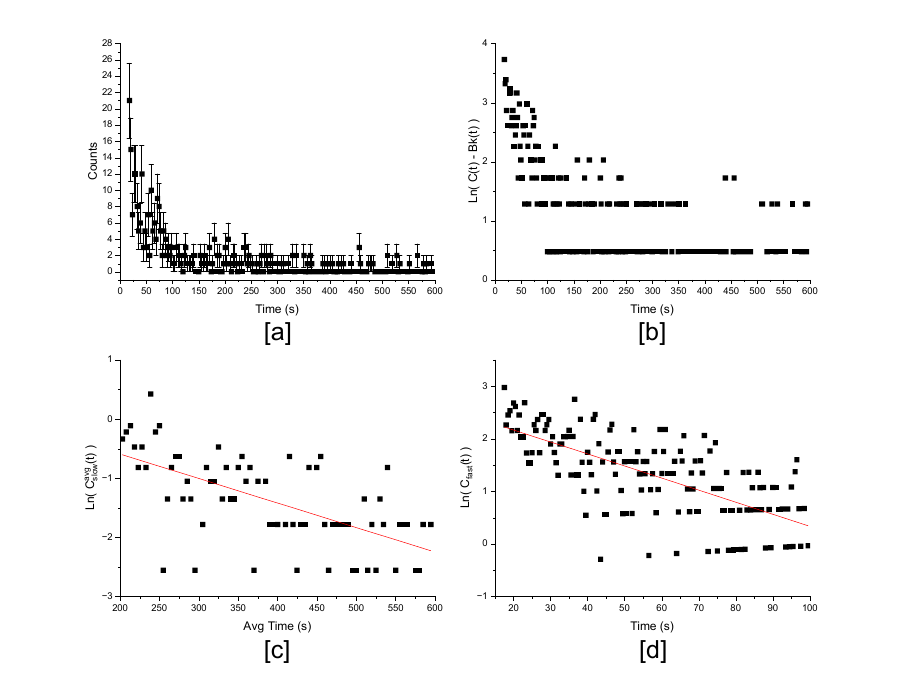}
\caption{Sample data and analysis graphs for the decay of silver experiment. a) Shows the total counts $C^{\rm T}(t)$ versus time $t$ for experimental silver data. The sample counts start at 17.5 seconds because the students started measuring counts before they placed the sample below the Geiger Counter. b) Shows $\ln[C^{\rm T}(t)-C^{\rm B}]$, where $C^{\rm T}(t)$ is the total counts and $C^{\rm B}$ is the average number of background counts. The knee is estimated to be at 100 seconds by observing the minimum of the residuals of a linear fit to this plot. c) Shows the linearized counts of $^{108}$Ag 100 s after the knee, where presumably all of the shorter half-life $^{110}$Ag has decayed and only $C^{\rm L}(t)$ is left. The fit parameters of the line give $\lambda= 0.00414\pm 0.00061 s^{-1}$ and $\ln[C_0^{L}]=0.237\pm 0.25$, the former of which corresponds to $t_{1/2}^{\rm L}=167\pm24$~s. d) Using the parameters from the linear fit of $\ln[C^{\rm L}(T)]$, the $C^{\rm L}(t)$ counts are subtracted from the $C^{\rm T}(t)-C^{\rm B}$ giving only counts due to $C^{\rm S}(t)$. This graph shows the $\ln[C^{\rm S}(t)]$ versus $t$ during the first 100 seconds where these counts are greatest. The slope of this graph gives $\lambda^{\rm S} = 0.0229\pm 0.0019 \; {\rm s}^{-1}$, which corresponds to $t_{1/2}^{\rm S}=30\pm3$~s. An over estimate of $t_{1/2}$ persists due to the use of least-squares fitting as discussed in Sec.~\ref{sec:poisson}.}
\label{fig:DecaySilver4Panel}
\end{figure}

The two-atom decay is much more complicated than the single-atom decay. Figure \ref{fig:DecaySilver4Panel} shows typical first-experimental data for the simultaneous decay of $^{108}{\rm Ag}$ and $^{110}{\rm Ag}$. The maximum number of counts is smaller than students expect based on their simulations, but there is the added difficulty that there are many subjective choices that students must make in establishing their data taking and analysis protocol. Students often find creative ways to address these choices, including interesting ways to analyse data and improve their counting statistics beyond just modifying data acquisition or the physical apparatus. For example, to find the knee, many students notice that the knee occurs roughly where the residuals from a linear fit of $\ln[C^{\rm T}(t)-C^{\rm B}]$ reach a minimum. For the data shown in Fig. \ref{fig:DecaySilver4Panel}a and b, students found $t_{\rm knee}=100$~s. Some students choose to divide the datasets for the fast and slow decay at the time they choose for the knee $t_{\rm knee}$, while other students choose the range of data analyzed for the short and long decay to eliminate some specified amount of time before and after the knee (e.g., in Fig.~\ref{fig:DecaySilver4Panel} students choose to start their analysis 100s after $t_{\rm knee}$).

Additionally students invent approaches to dealing with few counts during the later part of the experiment. In Fig.~\ref{fig:DecaySilver4Panel}c, students averaged the count rate for 10 data points and associated that average count rate with the average time for those count rates in order to address the limited number of counts at the end of the experiment. For the short half-life isotope (see Fig.~\ref{fig:DecaySilver4Panel}d), students did not feel the need to average. Given the various approaches that students can take to analyzing their data, there are ample opportunities for the instructor to probe the opportunities, challenges, and validity of students' approach to their work. In the end, given their analysis approach, this group of students found $t_{1/2}^{\rm S}=30\pm3$~s as the recovered experimental half life of $ ^{110}{\rm Ag}$ and $t_{1/2}^{\rm L}=168\pm24$~s as the recovered experimental half-life of $ ^{108}{\rm Ag}$. 

\section{Conclusion}
One of our main curricular goals is to help students understand the relationship between theory, models, and physical systems. Therefore rather than framing experiments in terms of proving physics theories, we focus on modeling a physical system and asking the question of whether our mathematical model includes the relevant physics to appropriately represent the essential features of the system. We then encourage students to consider how we use use computational and experimental work to deepen our understanding of the physical system by making observations and predictions based on the mathematical model.  By comparing their results to a system with known physical constants and behavior, students have the opportunity to assess whether their understanding, implementation, and/or analysis are robust and the instructor serves as a coach to probe these questions with students.

Expanding on what students learn from the single-atom decay, the computational portion of the two-atom decay lab allows students to observe and understand the evolution of the individual isotopes in a way that is less challenging when looking at the model as a whole. Thus the computational models allows students to understand how various parameters in their more complicated mathematical model impact what they might observe in the actual experiment.
Further connecting parameters in the simulation to the experimental settings, allows students to make an explicit connection between the computational model, their equipment settings, and the aspects of the physical system they will study.

The labs we discuss in this paper provide rich opportunities for students to iterate between mathematical models, computation, and experimentation to provide a more realistic experience of the relationship between the three. Allowing computation to inform experimentation, experimentation to inform computation, and both together to inform the mathematical model, helps students have a better understanding of the scientific method and how the scientific community knows what it knows. We encourage developing lab courses where students have the opportunity to combine theory, computation, and experiment and allow multiple lab periods for students to circle back to previous work (i.e., theory and models) with the new information they learned. In this way, students have a more authentic discovery experience than what is typically provided in prescriptive labs and/or siloed experiences (e.g., computational or experimental courses without integration of these two approaches). 

\begin{acknowledgments}
We would like to thank Andy Popick and Claire Kelling in the Carleton College Mathematics and Statistics Department for helpful conversations, and Barry Costanzi in the Carleton College Physics and Astronomy Department for helpful feedback on this paper. 
\end{acknowledgments}

\end{document}